\documentclass[twocolumn]{aastex631}
\pdfoutput=1
\usepackage{amsmath}

\begin{document}

\title{On the Impacts of Halo Model Implementations in Sunyaev-Zeldovich Cross-Correlation Analyses}

\author[0000-0002-6612-2524]{Chad Popik}
\affiliation{Department of Astronomy, Cornell University, Ithaca, NY 14853, USA}

\author[0000-0001-5846-0411]{Nicholas Battaglia}
\affiliation{Department of Astronomy, Cornell University, Ithaca, NY 14853, USA}
\affiliation{Universit\'e Paris Cit\'e, CNRS, Astroparticule et Cosmologie, F-75013 Paris, France}

\author[0000-0002-1048-7970]{Aleksandra Kusiak}
\affiliation{Institute of Astronomy, University of Cambridge, Cambridge, CB3 0HA, UK}
\affiliation{Kavli Institute for Cosmology, University of Cambridge, Cambridge CB3 0HA, UK}

\author[0000-0003-4922-7401]{Boris Bolliet}
\affiliation{Kavli Institute for Cosmology, University of Cambridge, Cambridge CB3 0HA, UK}
\affiliation{DAMTP, Centre for Mathematical Sciences, Cambridge CB3 0WA, UK}

\author[0000-0002-9539-0835]{J. Colin Hill}
\affiliation{Department of Physics, Columbia University, New York, NY 10027, USA}

\begin{abstract}

Statistical studies of the circumgalactic medium (CGM) using Sunyaev-Zeldovich (SZ) observations offer a promising method of studying the gas properties of galaxies and the astrophysics that govern their evolution. Forward modeling profiles from theory and simulations allows them to be refined directly off of data, but there are currently significant differences between the thermal SZ (tSZ) observations of the CGM and the predicted tSZ signal. While these discrepancies could be inherent, they could also be the result of decisions in the forward modeling used to build statistical measures off of theory. In order to see effects of this, we compare an analysis utilizing halo occupancy distributions (HODs) implemented in halo models to simulate the galaxy distribution against a previous studies which weighted their results off of the CMASS galaxy sample, which contains nearly one million galaxies, mainly centrals of group sized halos, selected for relatively uniform stellar mass across redshifts between $0.4<z<0.7$. We review some of the implementation differences that can account for changes, such as miscentering, one-halo/two-halo cutoff radii, and mass ranges, all of which will need to be given the proper attention in future high-signal-to-noise studies. We find that our more thorough model predicts a signal $\sim 25\%$ stronger than the one from previous studies on the exact same sample, resulting in a $33\%$ improved fit for non-dust-contaminated angular scales. Additionally, we find that modifications that change the satellite fraction even by just a few percents, such as editing the halo mass range and certain HOD parameters, result in strong changes in the final signal. Although significant, this discrepancy from the modeling choices is not large enough to completely account for the existing disagreements between simulations and measurements.

\end{abstract}


\section{Introduction} \label{sec:intro}

Galaxy formation and evolution are subjects of major interest in modern-day astronomy and cosmology, with a multitude of open questions. Among them is the observed inefficiency of star formation \citep{Fukugita2004, Gallazzi2008, Federrath2015}, the missing baryon problem \citep{Fukugita1998, Fukugita2004, Cen2006, Bregman2007}, and the role that feedback from supernovae (SN) and active galactic nuclei (AGN) plays in their host galaxies \citep{DiMatteo2005, Springel2005, Scannapieco2008, Somerville2008, Marasco2015, Somerville2015, Agertz2016}. 

A vital component for studying these questions is the circumgalactic medium (CGM), the region of diffuse ionized gas that surrounds a galaxy \citep{Tumlinson2017}. This region is known to have significant interaction with galaxies in the form of turbulence flows and thermodynamic exchanges \citep{Steidel2010, Nielsen2015, Peeples2019}. The CGM has been observed directly through absorption lines \citep{Lanzetta1995, Chen1998, Tumlinson2011, Rudie2012, Tumlinson2013, Werk2014, Chen2018, Lan2018, Zahedy2019, Wilde2021}, and improvements in observation technology have led to studies on emission lines \citep{Borisova2016, Emonts2016, Ginolfi2017, Wisotzki2018, ArrigoniBattaia2019, Leclercq2020, Zabl2021}. Furthermore, there have been many studies using simulations to understand the astrophysics involved in the CGM \citep{Oppenheimer2008, Ford2013, Hummels2017, Suresh2017, Hummels2019}.

Within the past decade, large galaxy redshifts surveys such as the Sloan Digital Sky Survey's (SDSS) Baryon Oscillation Spectroscopic Survey (BOSS) \citep[e.g.,][]{Alam2017} paired with high resolution wide field observations such as the tSZ, lensing \citep[e.g.,][]{Madhavacheril2024}, and X-rays \citep[e.g.,][]{Oren2024} opened the door to statistical studies of galaxies properties. With datasets including hundreds of thousands or millions of galaxies, researchers can extract general properties about the CGM that surrounds galaxies which then can be used to guide theoretical modeling and simulations. A promising method to do this is cross-correlating a given observation with the galaxies in its field, gaining an average measurements across a survey which represents the properties of these galaxies in general. 

One such observation is the thermal Sunyaev-Zeldovich (tSZ) effect, a secondary anisotropy of the Cosmic Microwave Background (CMB) caused by the inverse-Compton scattering of CMB photons off of electrons in hot regions of space \citep{Sunyaev1970}  \citep{Sunyaev1972}. The tSZ effect manifests as temperature fluctuations on the order of $10^{-5}$, and can be described as follows:
\begin{equation} \label{eq:TtSZ}
    \frac{\Delta T(\nu)}{T_{\text{CMB}}} = f(\nu) y(\theta) \ ,
\end{equation}
where $\Delta T(\nu)$ is the temperature deviation caused by the tSZ effect at a frequency $\nu$, $T_{\text{CMB}} = 2.725$ K is the temperature of the CMB, and $f(\nu)=x\coth{(x/2)}-4$ is the spectral function where $x=\frac{h\nu}{k_B T_{\text{CMB}}}$, where $h$ and $k_B$ are Planck and Boltzmann constants, respectively. The final term is the Compton-$y$ parameter, the integral of the electron pressure over the line of sight, defined as the following as a function of the angular size $\theta$:
\begin{equation} \label{eq:ComptonY}
    y(\theta) = \frac{\sigma_T}{m_e c^2} \int_{\text{LOS}} P_e \Big(\sqrt{l^2 + d_A^2(z)|\theta|^2}\Big)dl \ ,
\end{equation}
where $\sigma_T$ is the Thompson cross-section, $m_e$ is the electron rest mass, $c$ is the speed of light, $P_e$ is the electron pressure, $l$ is the line of sight distance, $d_A$ is the angular diameter distance at redshift $z$.

Maps of the tSZ therefore can be used as a probe of thermal energy of the matter distribution in the universe, and statistically averaging these measurements with galaxies would then yield an average pressure profile for that catalog of galaxies. Such analyses have been carried out with many different experiments before \citep{Spacek2017, Vikram2017, Tanimura2020, Koukoufilippas2020, Meinke2021, Schaan2021, Pratt2021, Yan2021}, using one of two nominal methods. The first involves taking cuts of the tSZ map reprojected to center each galaxy in the survey, and stacking these to get a mean signal of the survey galaxies. The second is creating map of the galaxy overdensity and cross-correlating it with the tSZ map to get a galaxy-$y$ cross-spectrum, which corresponds to a survey averaged tSZ profile (but in Fourier space). Both methods can be used to study the strength and shape of the tSZ signal, but each analysis has its own associated methodologies and systematics which must be accounted for when attempting comparisons between their results.

The goal of making such measurements is to then be able to extract astrophysical and potentially cosmological information out of them, by forward modelling analytic/semi-analytic models to produce a predicted signal that can then be fit to the results from data. To generate a statistical measurement from astrophysical modelling, it is also necessary to have a solid understanding of the underlying galaxy sample and its proper representation within the modeling, such as the distribution over redshift and halo mass. From the $\Lambda$CDM model of cosmology, galaxies and galaxy clusters are seeded within dark matter halos, and so these analyses often employ halo models \cite{Cooray2002, vandenBosch2013, More2015}, which calculate the distribution of matter in the universe and the evolution of structure using dark matter halos as the building blocks of large scale structure \cite{Vikram2017, Pandey2022, Koukoufilippas2020}. The distribution of galaxies can then be extrapolated from the halo model distributions, in a way adapted to the galaxy survey being analysed. 

Using emulators built off simulations in place of (semi)analytic models offers an opportunity to directly connect real data to the implementation of astrophysics in simulations, and possibly constrain the strength of thermodyanmic effects. Previous attempts to perform this have found that the predicted modelling from simulations significantly underestimates the signal in data \citep{Amodeo2021}. However, as noted earlier, there is significant variation between forward modeling procedures, which could introduce systematics and alterations in the results that would be an effect of methodology, instead of inherent different between simulations and measurements. In \citet{Moser2021} they investigated the importance of modeling the galaxy sample, such as different sample selections in the simulation and different fitting models. \citet{Moser2023} continued on this work, looking into systematics in the forward modeling process and finding minimal changes in altering the projection of 3D pressure profiles to observable signals. This study seeks to extend this analyses to the implementation halo models, specifically their use of Halo Occupancy Distributions (HODs) \cite{Zheng2005, Zehavi2011, vandenBosch2013, More2015} to populate dark matter halos with central and satellite galaxies. This is a process used in the modelling of galaxy-$y$ cross-spectra analyses, but not in stacking analyses such the previous two papers and \citet{Amodeo2021}, and therefore we compare the results of this analyses on the same data but using HOD modelling to see if the difference in modelling could be responsible for the discrepancies between measurements and simulations. Additionally, we also examine the effect of miscentering, which has been utilized in lensing analyses but has not yet been examined for tSZ cross-correlation studies \citep{McClintock2019}.

This paper is organized as follows: In Section~\ref{sec:methods}, we detail the methodology of forwarding a theoretical pressure profile to an observable average tSZ signal using a halo model (with HOD) to recreate the statistical measure. The starting profiles and projection functions used in~\ref{sec:Profiles} and~\ref{sec:Projection}, respectively, we seek to keep consistent between the two analyses, highlighting the differences in methodology arisen form the HOD and halo model details in~\ref{sec:HODs} and~\ref{sec:Fourier}, respectively. In Section~\ref{sec:results} we examine the effect of the HOD as well the effects of miscentering. Section~\ref{sec:HODResults} explores the HOD parameters we used and their impact on the tSZ signal. We then discuss our conclusions and the implications our findings have on future analyses of upcoming data in Section~\ref{sec:conclusion}.

\section{Theory and Methodology} \label{sec:methods}

The SZ measurements used throughout this work by \citet{Schaan2021} use the CMASS sample which is characterized as Luminous Red Galaxies (LRGs, \cite{Padmanabhan2007}), which are mainly centrally located in group-sized halos of masses $M_{\text{h}}<10^{14} M_\odot$. Galaxies in this survey occupy a redshift range of $0.4 < z < 0.7$ with a median value of $z=0.55$, and stellar mass and halo mass ranges of $10.71 < \log_{10}{(M_\star/M_\odot)} < 11.72$ and $12.12 < \log_{10}{(M_h/M_\odot)} < 13.98$, respectively. We will compare with the models from \citet{Amodeo2021} and \citet{Moser2021} that were used to analyze the SZ measurements from \citet{Schaan2021}. 

These models take a weighted average of 3D pressure profiles, as described in more detail in \ref{sec:Profiles}. The results is a matched stack that is projected to a observable signal at 150 GHz using the methodology in \ref{sec:Projection}. In our comparison, we used the same methodology to generate halo mass dependent 3D pressure profiles and to project 3D profiles to 2D observables, but take a HOD focused approach to generate the sample mass-average 3D pressure profile corresponding to the matched stack.

\subsection{Pressure and Density Models} \label{sec:Profiles}

We use a generalized Navarro-Frenk-White profile (GNFW) to describe the small-scale mass distribution of halos and galaxies, proposed by \citet{Zhao1996} based on the NFW profile \citep{Navarro1997}:
\begin{equation} \label{eq:GNFW}
    \rho(x) = \frac{\rho_0}{x^{\gamma} \big[1+x^{\alpha}\big]^{[\beta-\gamma]/\alpha}},
\end{equation}
where $x$ is a unitless scale defined as $x = \frac{r}{R_s}$ with $R_s$ as some scale radius, the central pressure $\rho_0$ is used as a normalization factor, and the rest are free parameters where $\gamma=1, \alpha=1, \beta=3$ sets a standard NFW profile. In \citet{Kou2023}, the authors fix $\alpha=\gamma=1$ while varying $\beta$ for the matter and satellite galaxy profiles as a fit parameter ($\beta_m, \beta_s$) along with the HOD parameters; in our study, we use their values for $\beta_s$, but use default values of $\beta_m=3$ to keep consistency with \citet{Moser2021}.

Halo models utilize a pressure profile, constructed from theory or simulations, to simulate the Compton-$y$ signal. For this study, both the real space and Fourier space predictions utilized the analytic model are detailed in \citet{Battaglia2012a}, which describes the electron pressure of a galaxy's CGM as a function mass, redshift, and distance to its center in the unitless scale $x$:
\begin{equation} \label{eq:Pe(x)} \begin{split}
    P_{e}(x; z, M_h) & = X_e \frac{\Omega_b}{\Omega_m} P_{200, c} (z, M_h) \ P_0(z, M_h) \\ 
    & \bigg[\frac{x}{x_c(z, M_h)}\bigg]^{\gamma}  \Bigg[1+\bigg[\frac{x}{x_c(z, M_h)}\bigg]^{\alpha} \Bigg]^{-\beta(z, M_h)} ,
\end{split} \end{equation}
where in this case the scale radius used to calculate $x=r/R_s$ is $R_{200c} (z, M_h)$, the radius of a sphere, centered around a halo of mass $M_h$, with density 200 times the critical density of the universe at redshift $z$. Additionally, $P_{200, c}(z, M_h) =  \frac{G M_{200, c}(z, M_h) [200\rho_{c}(z)]}{2 R_{200, c}(z, M_h)}$ is the pressure of a sphere with 200 times the critical density, $X_e=\frac{2(X_H+1)}{5X_H+3}$ is the electron fraction, and with fixed values $X_H=0.76, \Omega_b=0.044, \Omega_m=0.25, \gamma=-0.3, \alpha=1$. The parameters $P_0, x_c, \beta$ are fit to a power law $A(z, M_h) = A_0^x \Big[\frac{M_{200, c}(z, M_h)}{10^{14} M_\odot} \Big]^{\alpha_m^x} [1+z]^{\alpha_z^x}$ as described in \citet{Battaglia2012a}, from which we use their parameters.

\subsection{Halo Occupancy Distributions} \label{sec:HODs}

With identical pressure profiles, the goal of this study is now to compare the methods used to model the galaxy sample to obtain an average pressure profile. In \citet{Moser2021}, their sample is divided into four bins of equal size in stellar mass (which is given in the CMASS data) and assigned weights such that $\sum_i w_i \Delta m_{*, i} = 1$, where $\Delta m_{*, i}$ are the size of the stellar mass bins. This study then used the method described in \citep{Kravtsov2018} to obtain halo masses from galaxy stellar masses, and uses those and the weights to take a weighted average of halo mass of the pressure profile. 

HODs are used to describe how dark matter halos are populated with central galaxies, which reside at the center of their host halos, and satellite galaxies, which orbit around the centrals within the halo. The functional functional form of an HOD describes the average number of central and satellite galaxies as a function of host halo mass, usually prescribed to have a maximum of one central galaxy and to start adding satellites only after a halo is more like to host a central than not \citet{Zheng2005}. This framework is modified for each survey to account for the dependence on stellar mass, redshift, and survey incompleteness, resulting in a myriad of different HOD models between datasets and analyses.

In our analysis, we primarily used the HOD description in \citet{Kou2023} which follows the parameterization created for the CMASS galaxy sample in \citet{More2015}. The average number of centrals in a halo of mass $M_h$ is given by
\begin{equation} \label{eq:HODNc}
    \overline{N_{\text{c}}} (M_h) = \frac{f_{\text{inc}}(M_h)}{2} \bigg[ 1 + \text{erf} \Big[ \frac{\log M_h - \log M_{\text{min}}}{\sigma_{\log M_h}} \Big] \bigg] \ , 
\end{equation}
where erf is the error function, $\sigma_{\log{M_h}}$ controls the width or steepness of the transition between none and one central galaxy, and $M_{\text{min}}$ is the mass at which halos in the sample are more likely to have a central galaxy than not (as $\overline{N_{\text{c}}}(M_h \geq M_{\text{min}}) \geq \frac{f_\text{inc}}{2}$). $f_\text{inc} (M_h)$ is the incompleteness fraction of the galaxy survey, which characterizes the noise of the incompleteness as a number between 0 and 1, and is given by
\begin{equation} \label{eq:HODf_inc}
    f_{\text{inc}} (M_h) = \max{(0, \min{(1, 1+\alpha_{\text{inc}}[\log{M_h}-\log{M_{\text{inc}}}]})} ,
\end{equation}
where $\alpha_{\text{inc}}, M_{\text{inc}}$ are parameters used in to quantify the incompleteness in CMASS studies \citep{More2015} \citep{Kou2023}.

The mean number of satellite galaxies for a given halo mass $M_h$ is given by
\begin{equation} \label{eq:HODNs}
    \overline{N_{\text{s}}} (M_h) = \overline{N_{\text{c}}} (M_h) \bigg[ \frac{M_h-M_0}{M_1} \bigg]^{\alpha_s} \mathbb{H}(M_h > M_0),
\end{equation}
where $\mathbb{H}$ is the heaviside function,  $M_0$ is the minimum mass required for halos to contain satellite galaxies, $M_1$ is a characteristic mass which controls the number of satellites galaxies in massive halos, and $\alpha_s$ is the index of the power law that defines the satellite galaxy number. \citet{Kou2023} follows previous other studies using a fixed $\alpha_s = 1$ \citep{Zheng2005}, \cite{More2015}; additionally, they set $M_0 = M_\text{min}$, effectively stating that halos can start containing satellite galaxies at the same point as centrals.

Values for these parameters were found in \citet{Kou2023} through fitting their halo model against the angular auto-spectra of CMASS galaxies in BOSS DR12 \citep{Alam2015} and \emph{Planck} lensing and tSZ maps \citep{Planck2020, Planck2016}. The selected galaxy sample had redshift range of $0.47 < z < 0.59$ with a median value of $z=0.53$, and divided into subsamples using stellar mass thresholds for, whose values were obtained using the Portsmouth stellar population synthesis code \citep{Maraston2013}. We used the subsample corresponding to the lowest stellar mass threshold ($\log M_*/M_\odot > 11.1$, "A") and most galaxies (473,596) to most closely resemble the sample of galaxies used in \citet{Moser2021}.

\subsection{Power Spectra}

To probe the pressure profiles of a given galaxy sample, we calculate the galaxy Compton-$y$ cross-power spectrum $P_{gy}(k)$, which can be written as the sum of a one-halo and two-halo component:
\begin{equation} \label{eq:P1h+P2h}
    P_{gy}(k) = P_{gy}^{1h}(k) + P_{gy}^{2h}(k) \ ,
\end{equation}
where these terms are calculated separately in the halo model following the method in \citet{vandenBosch2013}. The one-halo component $P_{gy}^{1h}(k)$ comes from the correlation of galaxies and the pressure of their host halo, therefore dominating on scales of the halo radius and smaller, and it calculated as follows:
\begin{equation} \label{eq:Pgy1h} \begin{split}
    P^{1h}_{gy} (k) = \iint &  y (k; z, M_h) \mathcal{H} (k; z, M_h) n_h(z, M_h) dM_h dz \ ,
\end{split} \end{equation}
where $n_h(z, M_h)$ is the halo mass fraction, the number density of halos with mass $M_h$ at redshift $z$, and $y(k; z, M_h)$ is the the Compton-$y$ profile in Fourier space:
\begin{equation} \begin{split} \label{eq:ComptonYk}
    y(k; z, M_h) = \frac{\sigma_T}{m_e c^2}  \frac{4 \pi R_{200,c}^3 (z, M_h) [1+z]^2}{H(z)} & P_e(k; z, M_h) \\ & ,
\end{split} \end{equation}
where $H(z)$ is the Hubble function, and $P_e(k; z, M_h)$ is the electron pressure profile defined in Eq. \eqref{eq:Pe(x)} in Fourier space.

$\mathcal{H} (k; z, M_h)$ is the HOD modeled survey fraction of galaxies (both central and satellite) in halos with mass $M_h$ at redshift $z$:
\begin{equation} \begin{split} \label{eq:W_HOD}
    \mathcal{H} (k; z, M_h) = \frac{1}{n_{g} (z)} \Big[ & u_c (k; z, M_h) \overline{N_c} (M_h) \\ & + u_s (k; z, M_h) \overline{N_s} (M_h) \Big] \ ,
\end{split} \end{equation}
where $\overline{N_c}$ and $\overline{N_s}$ are defined in the previous section and $n_g(z)$ is the total galaxy number density calculated as follows:
\begin{equation} \label{eq:n_g}
    n_g (z) = \int \big[[\overline{N_c}(M_h) + \overline{N_s}(M_h)] \ n_h(z, M_h) \big] dM_h \ ,
\end{equation}
and $u_c(k;z, M_h)$ and $u_s(k;z, M_h)$ are the Fourier transformed density profiles for central and satellite galaxies, included to describe the distribution of galaxies within halos. In real space, $u_c$ is set to a Dirac delta at $x=0$ (which transform to $u_c(k;z,M_h)=1$ in harmonic space), equivalent to placing all centrals at the center of halos. Meanwhile, satellite galaxies are distributed around the center of their halos following a GNFW profile described in Eq. \ref{eq:GNFW} with $\beta$ being a parameter of the HOD model.

The two-halo component $P_{gy}^{2h}(k)$ comes from the contribution of galaxies in other halos and overtakes the rapidly decreasing one-halo component on larger scales, and is defined as follows:
\begin{equation} \label{eq:Pgy2h} \begin{split}
    P^{2h}_{gy} (k) = \int & P_\text{lin}(k; z) \\ & \bigg[  \int y(k; z, M_h)
    b_h(z, M_h) n_h(z, M_h) dM_h \\ & \int \mathcal{H}(k; z, M_h)
    b_h(z, M_h) n_h(z, M_h) dM_h  \bigg] dz
\end{split} \end{equation}
where $b_h(z, M_h)$ is the linear halo bias. We only consider the one-halo term throughout the study, using a constant two-halo term profile from \citet{Amodeo2021}.

\subsection{Halo Model Implementation} \label{sec:Fourier}

The survey averaged thermal pressure profile can be obtained by taking a weighted integral over the halo mass and redshift range of the survey in harmonic space:
\begin{equation} \begin{split} \label{eq:PthAv}
    \overline{P_\text{th}}(k) = \frac{ \iint  P_\text{th}(k;z, M_h) \mathcal{H} (k; z, M_h) n_h(z, M_h)  dM_h dz}{ \iint \mathcal{H} (k; z, M_h)  n_h(z, M_h) dM_h dz} ,
\end{split} \end{equation}
where the combined term $\mathcal{H} (k; z, M_h) n_h(z, M_h)$ acts as a galaxy weight modeling the distribution of the survey. 

It follows then that we can calculate the average $y$ profile in the same manner as Eq. \eqref{eq:PthAv} from the one-halo cross-spectra:
\begin{equation}
    \overline{y}(k) = \frac{P^{1h}_{gy} (k)} {\iint \big[ \mathcal{H} (k; z, M_h)  \ n_h(z, M_h) \big] dM_h dz} \ .
\end{equation}
Then, by dividing out the prefactor terms in Eq. \eqref{eq:ComptonYk}, we can arrive at an average pressure profile $\overline{P_\text{th}}(k)$, which we take an inverse Fourier transform to bring back to real space. We use the surveyed averaged values for these prefactor terms, obtained through the same weighted integral in Eq. \eqref{eq:PthAv}, to avoid reintroducing mass and redshift dependencies back into the survey averaged term. This allows us to directly compare the results of a real space model, which is unable to use HOD to model the galaxy distribution.

We used the halo model code \verb|hmvec|\footnote[1]{\url{https://github.com/simonsobs/hmvec}} to perform the calculations of required cosmological quantities (such as $H(z), n_h(z, M_h), R_{200, c}(z, M_h)$), where we used cosmological values of $h=0.673, \Omega_m = 0.314 = 1-\Omega_\Lambda, \Omega_b=0.0491$. \verb|hmvec|\ is also used to integrate over halo mass and redshift ranges to obtain $P_{gy}^{1h}(k)$, using the composite trapezoidal integration function \verb|numpy.trapz|, and to perform the necessary Fourier transforms to obtain $P_e(k;z,M_h), u_s(k; z, M_h)$ ,using the one-dimensional discrete fast Fourier transform function \verb|numpy.fft.rfft|, which is implemented as follows:
\begin{equation} \label{eq:FFT}
    P_e(k;z, M_h) = -\frac{\Delta_x}{k_t} \sum_{x=0}^{N_x-1} P_e(x;z, M_h) x e^{\frac{-2\pi i x k_t}{N_x}},
\end{equation}
where $\Delta_x = \frac{x_\text{max}-x_\text{min}}{N_x} = \frac{x_\text{max}-x_\text{max}/N_x}{N_x}$, and $N_x, x_\text{max}$ are defined as an input of the halo model. We convert from a dimensionless wavenumber $k_t$ corresponding to $x$ to a physical wavenumber $k=k_t/R_{200, c}(z,M_h)/[1+z]$ corresponding to $r$, and then interpolate the $P_e(k;z, M_h)$ arrays for all values of $z$ and $M_h$ onto the same array of $k$ values set by the model. To transoform back into real space, we used the inverse function \verb|numpy.fft.irfft|, and a survey averaged $\overline{k_t}$ is calculated through the same weighted integral in Eq. \eqref{eq:PthAv}, to once again avoid reintroducing mass and redshift dependencies.

Both \citet{Moser2021} and \verb|hmvec| use the general form of the profiles detailed Eqs. \eqref{eq:GNFW} and \eqref{eq:Pe(x)}, but there are distinct differences in the usage of mass and redshift values. As opposed to the virial halo mass $M_\text{vir}$ in \citet{Moser2021}, our halo model uses a mass definition of $M_{200c}$, the mass of a sphere centered on a halo with a density 200 times the critical density. This value is used to calculate the profiles $\rho(x)$ and $P_e(x;z,M_h)$, as well as the parameter within them, such as $R_{200, c}$ (and therefore $x$), $P_{200, c}$, $P_0$, $x_c$, and $\beta$. For this analyses, we edited our halo model to just use the halo mass. Furthermore, in \citet{Moser2021} the $P_{200, c}$ and $\beta$ values in the pressure profile were fixed, while the parameter $x_c(z, M_h)$ is determined from the mean redshift and halo mass, leaving only $R_{200, c}$ to change over the integration. This is different than our halo model, which varies all parameters through mass and redshift, and therefore finds an average pressure profile instead of the pressure profile at average values. To ensure that these misalignments are not creating artificial differences in our final results, we customized our halo model to match the mass and redshift values and treatment of parameters as in \citet{Moser2021}, and therefore ensure that the matter and pressure profiles being used for both methods were identical.

We further checked our calculations in sections 2.1-2.3 against another halo model \verb|class_sz|\footnote[2]{\url{https://github.com/CLASS-SZ}} \cite{Bolliet2023}, finding consistent values for the galaxy-$y$ cross-spectra when using the same profiles in \eqref{eq:Pe(x)} and the same HOD parameterization in Eqs. \eqref{eq:HODNc} and \eqref{eq:HODNs}.

\subsection{Projected tSZ Signal} \label{sec:Projection}

In \citep{Moser2021}, the authors projected the modeled profiles onto the sky to forecast direct observations of the tSZ signal stacked on galaxies. Following their methods, we use the  \verb|mop-c-gt|\footnote[2]{\url{https://github.com/samodeo/Mop-c-GT}} as implemented in \verb|SOLikeT|\footnote[2]{\url{https://github.com/simonsobs/SOLikeT}}, a likelihood package designed for use with the Simons Observatory. The projected average thermal pressure profile is calculated by integrating over the line of sight distance $l$:
\begin{equation} \label{eq:Project}
    \overline{P_\text{th}} ^{2D}(R) = 2 \int_0^\infty \overline{P_\text{th}}\Big( \sqrt{l^2 + R^2}\Big) dl \ ,
\end{equation}
where $R(\theta) = \theta d_A(\overline{z})$ is the projected size of the sky using a survey averaged redshift. As detailed in \citep{Moser2023}, the actual values for the step size and range of the integral can be significant to the end results. We found that integrating through $l=10^{-4}-50$ Mpc is sufficient to capture all of the signal over all relevant values of $R$, using 50 logarithmically space bins. Additionally, as this and following integrals will cover values of $r(R, l)$ that are below the lowest value of $r$ returned by the halo model, we take advantage of the plateau that occurs in the pressure profile at about $10^{-3}$ Mpc to extrapolate that value of $\overline{P_\text{th}}(r)$ for any $r$ values outside the normal range. 

Combining this with eqs \eqref{eq:TtSZ} and \eqref{eq:ComptonY}, the final modeled tSZ signal is: 
\begin{equation}
    \Delta T_\text{tSZ}(\theta; \nu) = T_\text{CMB} \ f(\nu) \ \frac{\sigma_T X_e}{m_ec^2} \ \overline{P_\text{th}} ^{2D}(\theta) \ .
\end{equation}
To account for the effects of the instrument, the signal is convolved with the beam of the instrument and is filtered using aperture photometry:
\begin{equation}
    \tilde{T}_{tSZ} (\theta_d) = \int [T_{tSZ}(\theta) \ \circledast \ B(\theta)]  \ W_{\theta_d}(\theta) \ d^2 \theta \ ,
\end{equation}
where $B(\theta)$ is the instrument beam convolved using a Hankel transform, $\theta_d$ is the aperture radius, and $W_{\theta_d}$ is the aperture filter which is 1 when $\theta<\theta_d$, -1 when $\theta_d \leq\theta\leq\sqrt{2}\theta_d$, and 0 otherwise.

\begin{figure*}[hbt!] \plotone{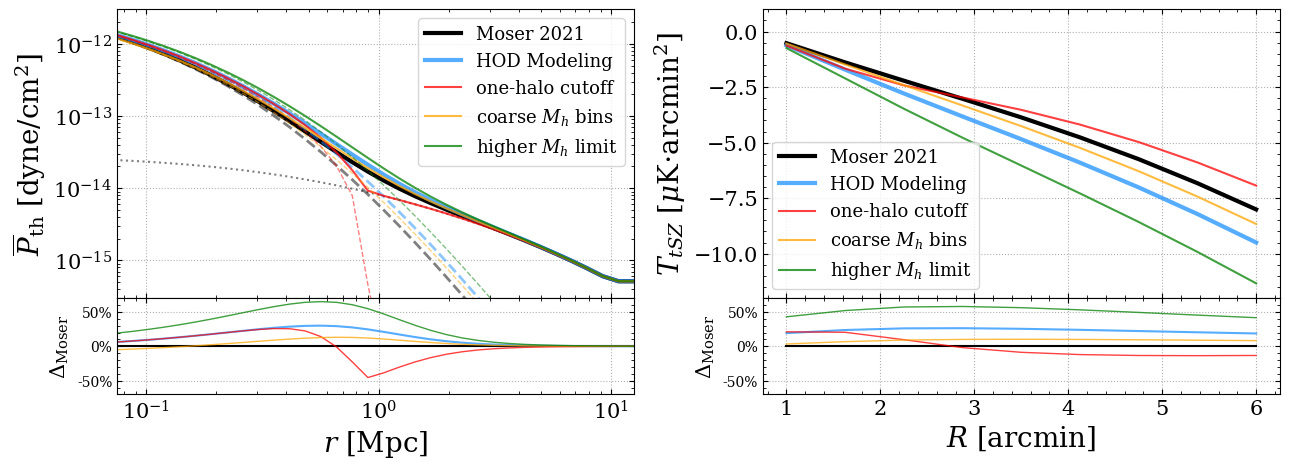} 
  \caption{Modeled survey averaged thermal pressure profile (left) and corresponding projected tSZ signal (right) using the methodology from \citet{Moser2021} (black) and the HOD modeling of this study (blue), with \% differences from \citet{Moser2021} shown below. Additionally shown are the results when incorrectly including inconsistencies such as a cutoff radius to the one halo term (red), very coarse bins for halo mass (yellow), and a higher upper limit of halo mass range (green), all of which cause significant deviations. Dashed and dotted lines represent the one-halo and two-halo component, respectively, where the latter is held constant throughout this study.}
\label{fig:corrections}  \end{figure*}

\subsection{Miscentering} \label{sec:Miscentering}

In a typical halo model, central galaxies are placed directly in the center of a halo, and therefore aligned with the pressure profile of that halo, while satellite galaxies are equally distributed at a distance from the halo center. This is based off the assumption that halo centers are properly identified and aligned with the galaxies within them; however, if a significant portion of galaxies are misaligned within their halo's pressure profile, this would dampen the cross-correlation signal between galaxies and pressure. In the mass averaged pressure profiles, this would manifest as a decrease in the signal at lower scales, with a boost at some scale corresponding to description of the miscentered galaxies. To incorporate this effect into our analysis, we follow the formalism used in \citet{McClintock2019} which follows \citet{Melchior2017} to model miscentered galaxies in calculating the lensing signal from clusters, replaced the lensing signal $\Delta \Sigma$ with the projected average thermal pressure detailed in Eq. \eqref{eq:Project}:
\begin{equation} \label{eq:PthwMis}
    \overline{P_\text{th}}_\text{, mis}^{2D} (R) = [1-f_\text{mis}]\overline{P_\text{th}}^{2D}(R) + f_\text{mis}\overline{P_\text{th}}_\text{, mis}^{2D}(R) \ ,
\end{equation}
where $f_\text{mis}$ is the fraction of miscentered galaxies and $\overline{P_\text{th}}_\text{, mis}^{2D}(R)$ is the averaged profile for miscentered galaxies, obtained by taking a weighted integral over the range of possible miscentering values:
\begin{equation} \label{eq:PthMis}
    \overline{P_\text{th}}_\text{, mis}^{2D}(R) = \int \overline{P_\text{th}}_\text{, mis}^{2D} (R | R_\text{mis}) \ p (R_\text{mis}) \ dR_\text{mis} \ ,
\end{equation}
where we integrate over $R_\text{mis} = 2\times10^{-4}-20$ Mpc with 100 logarithmically spaced bins, $\overline{P_\text{th}}_\text{, mis}^{2D} (R | R_\text{mis})$ is the profile for a miscentering value of $R_\text{mis}$, obtained by taking the azimuthal average of the surface thermal pressure:
\begin{equation} \begin{split} \label{eq:PthMisRmis}
    \overline{P_\text{th}}_\text{, mis}^{2D} & (R | R_\text{mis}) = \frac{1}{2\pi}\int_0^{2\pi} \overline{P_\text{th}}^{2D}\Big(R'(R, R_\text{mis}, \theta) \Big)d\theta \ ,
\end{split} \end{equation}
where $R'(R, R_\text{mis}, \theta) = \sqrt{R^2+R_\text{mis}^2+2RR_\text{mis}\cos\theta}$ using 50 linearly spaced bins, and $p (R_\text{mis})$ is the distribution of miscentering values of the model, which we set as a Gamma distribution following \citet{McClintock2019}:
\begin{equation} \label{eq:gamma}
    p(R_\text{mis}; \tau) = \frac{R_\text{mis}}{[\tau \overline{R_{200, c}}]^2} \exp \bigg(-\frac{R_\text{mis}}{\tau \overline{R_{200, c}}} \bigg) \ ,
\end{equation}
where $\tau = R_\text{mis} / \overline{R_{200, c}}$ sets the peak of the distribution as a function of some fractional miscentering value.

\section{Results} \label{sec:results}

We find that after several corrections and consistency edits, the two methods of generating theoretical predictions for the tSZ signal produce results similar in shape but noticeably different in scale. Our model generates a signal stronger than that from \citet{Moser2021} by about $\sim 20\%-25\%$, depending on the radius. While there exists some level of uncertainty as a result of small differences between the samples used in the two studies, this is a larger deviation than would be expected. 

\subsection{Implementation Effects} \label{sec:implementationeffects}

As shown in Figure \ref{fig:corrections}, there are several corrections that we have to make to the default halo model implementation to improve the consistency between our study and that of \citet{Moser2021}. These corrections wind up having moderate to significant effects on the final signal.

Even after ensuring the same thermal pressure profile for any given value of $r$, $z$, and $M_h$, there is still a significant dependence on the properties of the halo masses being used for averaging. Primarily, this is seen in the halo mass range: increasing the maximum halo mass includes more large halos with stronger pressure profiles, raising the average halo mass and therefore strengthening the pressure profile overall \citep{Moser2023}. As shown in Figure \ref{fig:corrections}, even a 50\% higher $M_h$ upper limit increases the difference from ~20\% to ~50\%. Additionally, changing these mass ranges affects the fraction of central/satellite galaxies within the survey, which are entirely dependent on the halo mass as shown in Eq. \ref{eq:HODNc} and \ref{eq:HODNs}, which alters the shape of the profile. To manage this, we restricted the mass range of our halo model to only integrate over the CMASS bins used for the real space average, halos masses between $12.12 < \log_{10} (M_h/M_\odot) <13.98$. Another effect is the choice of binning, both in number and spacing. Even with both models use logarithmically spaced bins, we found that changing our bin number by less than an order of magnitude can affect the signal by $10-20\%$.

It is not unusual for a halo model to differentiate between the one-halo and two-halo term by restricting their contributions to the scales where each term should be dominant. Our halo model adds a cutoff to the one-halo term in the pressure profile by applying a window function before performing the transform in Eq. \ref{eq:FFT}, resulting in the one-halo contribution to be zero at regions beyond $R_\text{vir}/R_{200, c}$. As can be seen in the dashed red line in Figure \ref{fig:corrections}, this procedure significantly changes the shape and scale of the average pressure profile, and as it is not performed in \citet{Moser2021}, we remove this step in our halo model.

\begin{figure*}[ht!] \plotone{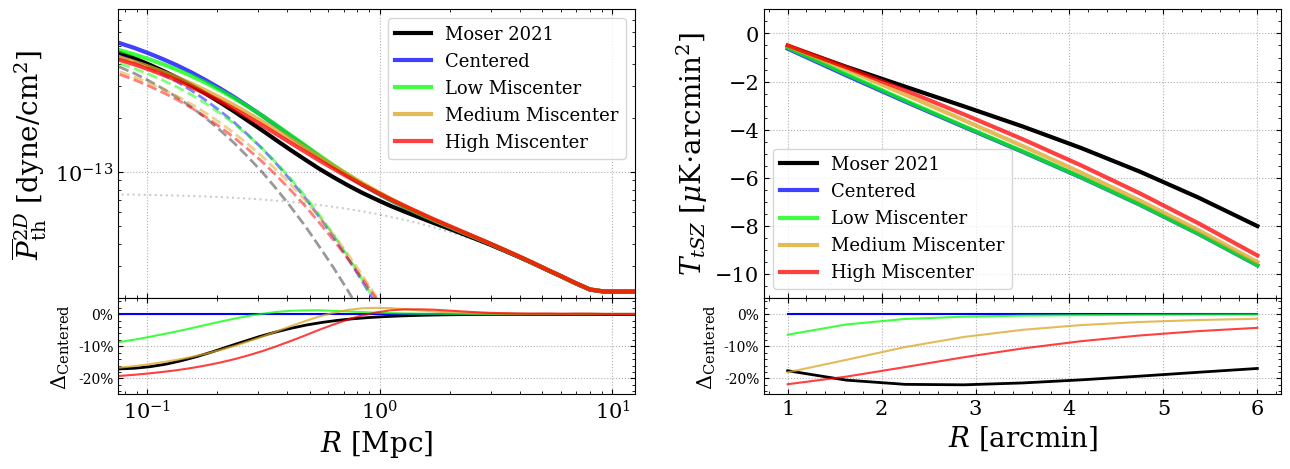}
\caption{Projected modeled survey averaged thermal pressure profiles (left) and corresponding observable tSZ signal (right) from \citet{Moser2021} (black), using our HOD modeling without miscentered central galaxies (blue), and with 25\% of central galaxies miscentered at small (green), moderate (yellow), and large (red) distances, with \% deviations from the centered model shown below. Miscentering is shown to dampen the signal more strongly for higher values of miscentering distance and at lower angular scales. Also included is the results from \citet{Moser2021} (black).}
\label{fig:Miscenter} \end{figure*}

Overall, the inclusion of an HOD when modeling the CMASS tSZ cross-correlation signal differs from the analyses in \citet{Amodeo2021,Moser2021,Moser2022} by $\sim 30\%$ depending on the angular scale. After accounting for different ways of modeling the mass distribution of the CMASS sample, it is the inclusion of satellites that has the largest impact on this tSZ signal because satellite galaxies populate more massive central galaxy/group with larger tSZ signals. The previous models would use the mass of the satellite for the tSZ signal, where instead it should be the satellite and its central galaxy/group. Here, the galaxy/group will dominate the signal given the steep mass scaling of $M^{5/3}$ of the tSZ signal, so at percent level accuracy the satellite contribution is negligible. This is explored more in the following subsection.

For miscentering, \citet{Zhang2019} showed that redMaPPer clusters have a miscentering fraction of about $f_\text{mis}\approx25\%$, a value we followed in our study. They based their values of $\tau$ using radius definition from redMaPPer that did not translate to our study, so we selected values of the combined term $\tau R_\gamma = 0.1, 0.3, 0.5$, chosen to show a low, moderate, and high level of miscentering. As shown in Figure \ref{fig:Miscenter}, our model acts as expected, weakening the pressure at low $R$ and strengthening it at high $R$, with higher $\tau R_\gamma$ or $f_\text{mis}$ values both increasing the degree of the effect; however, after transforming this to a final $T_{\text{tSZ}}$ signal, we see a weakening effect which is more potent at small scales, but we see no strengthening in the signal at any scale. We attribute this to the aperture filter applied in the transform which removes higher radii. We attempt to set a very low $\tau$ value to keep the profiles within the aperture, and see the increase at higher radii, but with such a small difference this strengthening effect is also very small to the point of negligible. The inclusion of satellites in the tSZ modeling is similar to that of miscentering, since the halo center will be that of the satellite which is off-centered from the peak on the pressure profile which is centered on the halo. In a way these two effects are degenerate with one another.

\subsection{HOD Parameters} \label{sec:HODResults}

\begin{figure*}[ht!] \plotone{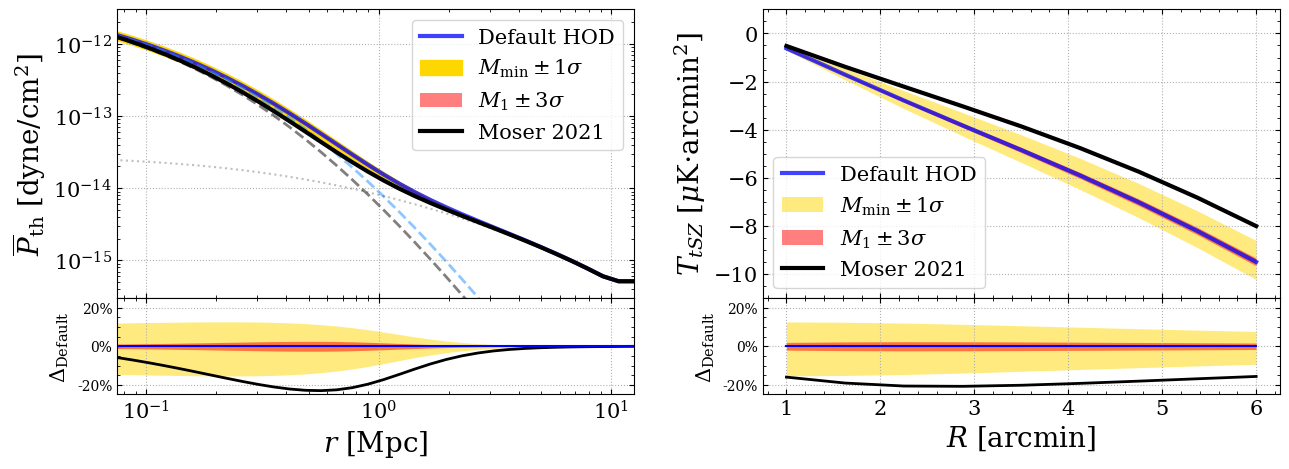}
\caption{PModeled survey averaged thermal pressure profiles (left) and corresponding observable tSZ signal (right) from the methodology of \citet{Moser2021} (black), using our HOD modeling with the default HOD parameters (blue), and showing regions of error resulting from $1\sigma$ deviations in the HOD parameter $M_\text{min}$ (yellow) and $3\sigma$ deviations in the HOD parameter $M_1$ (red), with \% deviations to default results shown below. While changes in $M_\text{min}$ have a clear effect, even unreasonably large changes in $M_1$ barely move the results as a result of nearly all satellite galaxies being cut off by the mass range of the halo model.}
\label{fig:parameters} \end{figure*}

Within the HOD model the different parameters have different physical meanings and influences on the final results. We discuss the meaning and impact of the different parameters below. 

The first parameter is the characteristic minimum mass of a halo containing a central galaxy and denoted as $M_{min}$, which is this study is also set to the value of $M_0$, the cutoff mass of a halo containing a satellite galaxy. Changing this parameter effectively changes the galaxies being selected: a very high value selects only galaxies with more massive halos, and a small value allows less massive halos (and hence less massive central galaxies) into the analysis. As a result, the value of this parameter has a strong effect on the average pressure profile, as the inclusion of less or more massive galaxies will yield a stronger or weaker pressure profile, respectively, just from the density and concentration of matter. It should be noted that this effect will also be dependent on the mass range being imposed on the analysis, as changing $M_{min}$ from a value of $10^{13} M_\odot$ to $10^{12} M_\odot$ will be less prevalent if only averaging starting with halos at $M_h\geq10^{12.95} M_\odot$ to match your survey. It would increase the number of galaxies in low mass halos towards the bounds, but would include no new galaxies in halos below the minimum mass in the actual average.

The other major parameter in $N_c$ is denoted as $\sigma_{\log{M}}$, which describes the steepness/speed of the transition from zero to one central galaxies. This parameter is occasionally seen in different forms (occasionally seen as just $\sigma/\sigma_M$), sometimes with a $\sqrt{2}$ directly outside it and sometimes fit as $\sigma^2$ instead, but always the denominator within the error function. A low value of this parameter yields a steeper transition at the cutoff mass $M_{min}$, while a high value creates a smooth transition around that point. Then, lowering the value of $\sigma_{\log{M}}$ spreads some of the halo mass right above the minimum mass to right below the minimum mass, effectively including more low masses and less higher masses, which has the effect of weakening the pressure profile. This parameter is therefore highly degenerate with $M_{min}$, though the profile is not as sensitive to changes in $\sigma_{\log{M}}$. Because of this, changing both parameters allows changing the shape of the mean number distribution while keeping the same mean mass in the analysis.

In most recent models, the term for central galaxies have also included an $f_{inc}$ term which represents the incompleteness fraction of the survey. This is a number between 0 and 1 determined by the survey specifics, and in the case of CMASS models, it is calculated with a function that includes two other parameters, $\alpha_{inc}$ and $M_{inc}$, to characterize the noise of the incompleteness. Varying this parameter causes small changes in the amplitude of the pressure profile without changes to the shape.

In the term for satellite galaxies, the major parameter of interest is $M_1$,  described as a characteristic mass of halos with satellite galaxies and controls the number of galaxies at high halo mass. Increasing this value raises the benchmark for satellite to exist in halos and therefore filters out galaxies in high mass halos, lowering the pressure profile. However, in our analysis, the value of $M_{min}$ is very close to the highest possible mass defined by our bins. While the $\sigma_m$ parameter in the $N_c$ term allows for some spread and therefore halos to exist below this value, in the $N_s$ equation this is a hard cutoff, and therefore practically zeroes the number of satellite galaxies in this survey regardless of $M_1$. This is shown in Figure \ref{fig:parameters}, where we see that $3\sigma$ error values on $M_1$ still have very little effect on the tSZ signal, and are dwarfed by the effect of $M_\text{min}$. To examine the effect of this parameter, we therefore had to artificially lower the value of $M_{min}$ into the range set by the mass bins.

This particular HOD also incorporated two parameters $\beta_m$ and $\beta_s$ which are used in the GNFW profiles for matter and satellite galaxies, respectively. We found that changing these parameters had no noticeable effect on the pressure profiles, even with $M_{min}$ shifted to artificially include more satellites.

\section{Discussions and Conclusions} \label{sec:conclusion}

\begin{figure}[ht!] \centering \includegraphics[width=0.47\textwidth]{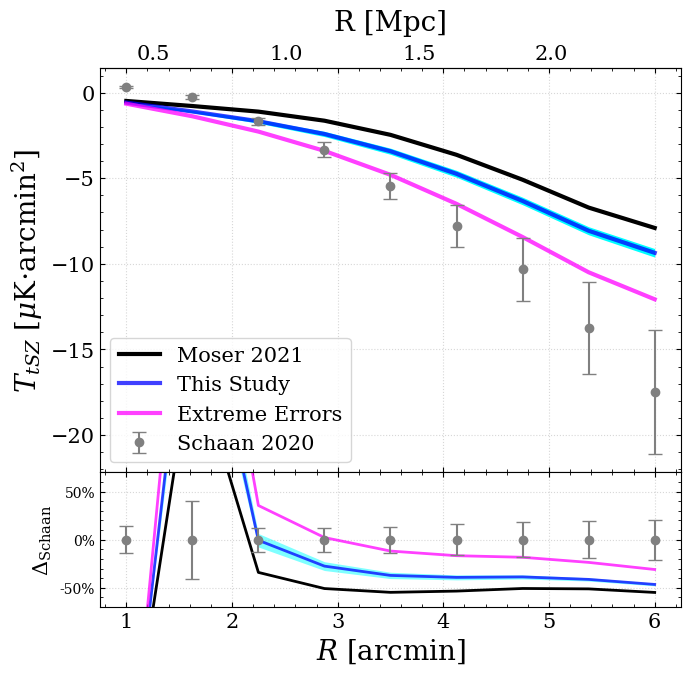}
\caption{Modeled observable tSZ signal from our HOD study (blue), with a $1\sigma$ confidence interval from the HOD parameter uncertainties, and from the methodology of \citet{Moser2021} (black), both containing an estimated dust profile from \citet{Amodeo2021}, compared to the ACT DR5 data results from \citet{Schaan2021} (grey), with \% deviations from the data shown below. While our study is unable to resolve the discrepancy between predicted and observed signals, the new modeling process yields a result that is more similar to the data than older results. Assigning a "worst case" scenario with several factors in this paper combined results in a signal that is much closer to data at larger angular scales at the expense of creating larger gaps at lower scales (magenta).} 
\label{fig:summary} \end{figure}

We find that differences in survey modeling methodology does manifest in significant differences in predicted tSZ signals. Our study using halo occupancy distributions produced profiles that are up to $25\%$ stronger than studies modeling the distribution of galaxies from stellar masses and without accounting for satellite galaxies. As shown in Figure \ref{fig:summary}, this shifts the predicted signal in the direction of the results from data and produces a better fit (without considering the angular scales below 2' which are heavily contaminated with dust), improving the $\chi^2/\text{DOF}$ by 33\%, from 3.88 to 2.60, and the probability to exceed value by 3576\%, from $3.1 \times 10^{-4}$ to 0.011. Despite the stronger signal we find in our study, our results are unable to explain the much larger discrepancies between simulations and observations, even accounting for reasonable errors in our models and systematics. 

Individual variations with a $1\sigma$ range of in the HOD parameters, which manifest as alterations in the fraction of satellite galaxies, could account for $10-20\%$ variations in the tSZ signal, but altogether the errors obtained from using the output of chains from the HOD parameter fitting process show tights constraints on the signal. Furthermore, there are unique systematics introduced with each method that will need to be accounted for on when comparing different studies and datasets. We also examine the effect of miscentering on observed tSZ signals, finding that it is capable of dampening the signal by equivalent amounts at low angular sizes, but is unable to strengthen the signal due to aperture photometry. We explore the results if several of the effects in this paper are combined, specifically a 50\% higher halo mass limit, a $+1\sigma$ error on $M_\text{min}$, and a $\tau \overline{R_{200, c}} = 0.5$ miscentering on 25\% of central galaxies. In this case, our predicted signal aligns much more closely with the data at larger angular scales, at the expense of increasing the gap at smaller angular scales. While it is unreasonable to believe that modeling errors would both be this extreme and all point in the direction of a stronger signal, this shows that there is weight to the idea that the disagreements between data and models are the result of several separate, smaller effects, and therefore continued refining of each step in the analysis pipeline is essential going forward.

\begin{table}[hbt!] \centering \begin{tabular}{|l|l|l|l|}
\hline
Effect               & $T_{tsz}(R<2')$ & $T_{tsz}(R>3')$ & $\Delta \%_\text{sat}$ \\ \hline
5 $M_h$ bins     & -15\%          & -10\%             & -1.7\%                   \\ \hline
+50\% max $M_h$  & +22\%          & +22\%            & +2.1\%                  \\ \hline
-50\% min $M_h$ & -2\%           & -1\%              & 0\%                   \\ \hline
one-halo cutoff      & -1\%           & -28\%             & N/A                     \\ \hline
                     &                &                   &                         \\ \hline
$\log M_\text{min} \pm 1\sigma$         & $\pm$13\%         & $\pm$9\%             & $\mp$1\%              \\ \hline
$\log M_1 \pm 1\sigma$         & $\pm$0.7\%        & $\pm$0.6\%           & $\pm$0.5\%           \\ \hline
                     &                &                   &                         \\ \hline
Low Miscent.     & -5\%           & 0\%               & N/A                     \\ \hline
Med. Miscent.  & -15\%          & -5\%              & N/A                     \\ \hline
High Miscent.    & -20\%          & -10\%             & N/A                     \\ \hline
\end{tabular}
\caption{Table of various implementation choices in our HOD forward modeling procedure discussed in this paper and their effect on the final tSZ signal, shown as \% differences from the final modeled signal of our study at small and large angular scales. Also shown is the changes in the percent of satellite galaxies in the modeled survey $\%_\text{sat}$ from the default value of 4.3\%. }
\label{tab:effects} \end{table}

\begin{figure*}[ht!] \plotone{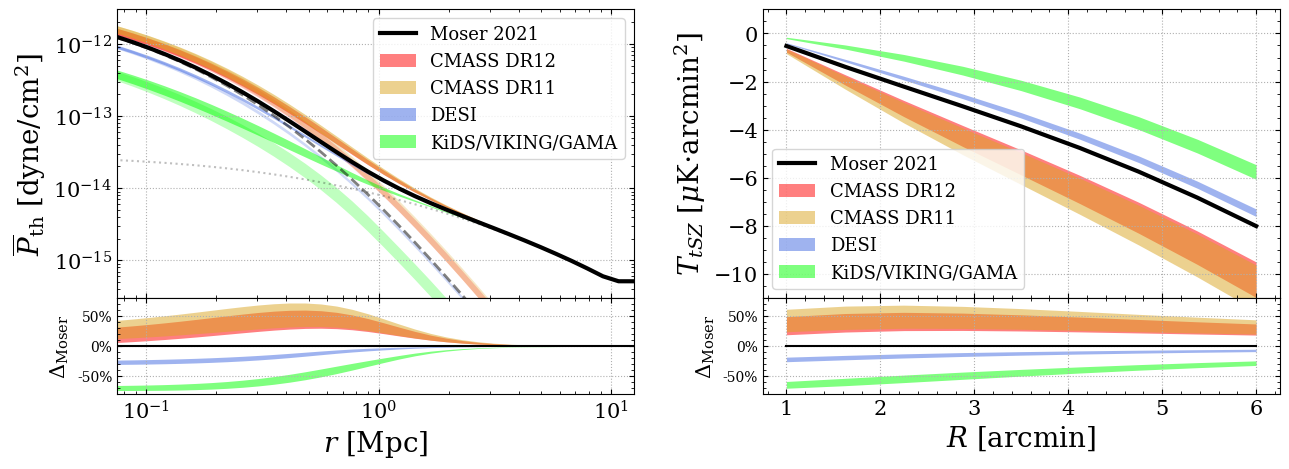}
 \caption{Average thermal pressure profiles and projected tSZ signal for several different HOD parameterizations, shown in bands to indicate the range of profiles for the subsets detailed in the studies, divided by stellar mass or redshift. \citet{Kou2023} (red) and \citet{More2015} (yellow) both model the CMASS galaxy sample examined in this study, while HODs for DESI LRGs \cite{Yuan2024} (blue) and KiDSxVIKING/GAMA \cite{Linke2022} (green) are included to show the larger difference between HOD models of different galaxy samples. The non-HOD model results \citet{Moser2021} show similar levels of deviation from HOD model results its modeled survey (CMASS) as completely different surveys, showing the significance of the modeling methodology on final profile results.}
\label{fig:HODPths} \end{figure*}

Recently, work by \citet{McCarthy2024} looked at the impact of satellites on the kSZ cross-correlation measurements \citep{Hadzhiyska2024} using numerical simulations. They found that including all satellite halos down to a given stellar mass threshold changed their predicted kSZ signal for CMASS galaxies up to 30\%. Similar to the tSZ result shown in this work, these kSZ changes are the result of satellite galaxies residing in more massive dark halos hence larger SZ signals. Interestingly the change induced by the inclusion of satellites in kSZ signal that \citet{McCarthy2024} found has a similar amplitude to what we found for tSZ signal. However, with simple mass scaling arguments for the kSZ ($\propto M$) and the tSZ ($\propto M^{5/3}$), one would expect that impact on the tSZ signal should be larger than the kSZ signal. One, essential, difference between these two analyses is that our HOD implementation models the selection of CMASS galaxies through the $M_0$ and $M_1$ parameters which account for the fact that CMASS galaxies were selected to be predominantly central galaxies and have a small satellite fraction for their stellar mass threshold according to clustering and weak-lensing measurements \citep{Ahn2014,Saito2016,Leauthaud2017,Kou2023}. Hence, the impact that the inclusion of satellites has on the tSZ signal in our methodology is not as large as if we were to include all galaxies (central and satellites) above a threshold mass disregarding how the CMASS galaxies were selected.

As upcoming observations continue to increase in scope and fidelity, the need for improved theoretical model of these future signals is made clear by this work. Next generation CMB experiments like the Simons Observatory \citep{Ade2019} and CMB-S4 \cite{Abazajian2016} will provide high fidelity and resolution measurements of the tSZ (and kSZ) by improving upon ACT, and current and upcoming galaxy surveys like the Dark Energy Spectroscopic Instrument (DESI) have a target density over ten times that of CMASS \citep{DESICollab2016}. As both will cover a larger fraction of the skym the overlap between the observations will also increase, and as the inherent noise is reduced by a factor of the square root of the number of galaxies in the study, this further improves the signal. Proper usage of HODs in halo models will be critical to quantifying galaxy properties of these future measurements.

We note that while the results should be broadly applicable, and our process is able to use any given HOD as shown in Figure \ref{fig:HODPths}, some of the analysis done in this project is specific to the HOD made for CMASS. Other HODs may include or omit other parameters that are explored in this project. For example, HODs being constructed for the DESI survey do not explore deviation from the standard NFW profile \citep{Yuan2022, Yuan2023, Yuan2024, Rocher2023}.

For this study, pressure profiles are obtained using the analytic model described in \citet{Battaglia2012a} (shown in Section \ref{sec:Profiles}) as is standard for halo model calculations. We note that the initial pressure profile parameters were fit to simulations clusters with halo mass of $M_{200,c} > 5 \times 10^{13}M_\odot$, which is above the mass range of CMASS halos. The fit to CMASS galaxy sample yielded a different parameterization that better fits the data. This fit is shown in \cite{Amodeo2021} to the \cite{Schaan2021} measurements to most some of the parameters of Eq. \ref{eq:Pe(x)}. Going forward such fits should use the forward model with an HOD outlined in this study. Multiple types of profiles can also be obtained through hydrodynamical simulations such as IllustrisTNG \citep{Pillepich2018, Nelson2018, Naiman2018, Marinacci2018, Nelson2019} and SIMBA \citep{Dave2019}, through stacking simulated galaxies or through emulators \cite{Moser2022, Hernandez-Martinez2024}. The predictions produced from such studies have similarly yielded a poor fit to observations, and have further shown that the modeled signal can be improved by increasing the strength of feedback in simulations \cite{Moser2022, Hadzhiyska2024}. The addition of the The CAMELS project \citep{VillaescusaNavarro2021, VillaescusaNavarro2022}, which has already shown encouraging results and continues to improve in detail and in size, is key for future comparison to simulations, where the parameters of sub-models which control feedback processes are varied.

We are grateful to Simone Ferraro, Emmanuel Schaan, and Sandy Yuan for their helpful feedback and useful discussions. NB acknowledges support from NASA grants 80NSSC18K0695 and 80NSSC22K0410. JCH acknowledges support from NSF grant AST-2108536, NASA grant 80NSSC22K0721 (ATP), the Sloan Foundation, and the Simons Foundation.

\bibliography{main.bib}{}
\bibliographystyle{aasjournal}

\end{document}